\documentclass{IOS-Book-Article}

\usepackage{times}
\normalfont
\usepackage[T1]{fontenc}
\usepackage{graphicx}
\usepackage{hyperref}
\usepackage{url}
\urldef{\mail}\path|m.baldi@univpm.it|

\begin{document}
\begin{frontmatter}                           

\title{LDPC codes in the McEliece cryptosystem: attacks and countermeasures}
\runningtitle{LDPC codes in the McEliece cryptosystem: attacks and countermeasures}

\author{\fnms{Marco} \snm{Baldi}
\thanks{The author is with the Department of Biomedical Engineering,
Electronics and Telecommunications, Polytechnic University of
Marche, Ancona, Italy; E-mail: \mail.}}
\runningauthor{Marco Baldi}
\address{Polytechnic University of Marche, Ancona, Italy}

\begin{abstract}
The McEliece cryptosystem is a public-key cryptosystem based on coding theory
that has successfully resisted cryptanalysis for thirty years.
The original version, based on Goppa codes, is able to guarantee a high level
of security, and is faster than competing solutions, like RSA.
Despite this, it has been rarely considered in practical applications, due to 
two major drawbacks: i) large size of the public key and ii) low transmission 
rate.
Several attempts have been made for overcoming such drawbacks, but the 
adoption of most families of codes has not been possible without compromising 
the system security.
Low-Density Parity-Check (LDPC) codes are state-of-art forward error 
correcting codes that permit to approach the Shannon limit while ensuring 
limited complexity.
Quasi-Cyclic (QC) LDPC codes are a particular class of LDPC codes, able 
to join low complexity encoding of QC codes with high-performing and 
low-complexity decoding techniques based on the belief propagation
principle.
In a previous work it has been proposed to adopt a particular family of QC-LDPC 
codes in the McEliece cryptosystem to reduce the key size and increase the 
transmission rate.
It has been shown that such variant is able to counter all the classic attacks, 
and also attacks that can compromise the security of previous LDPC-based 
versions.
Recently, however, new attacks have been found that are able to exploit
a flaw in the transformation from the private key to the public one.
Such attacks can be effectively countered by changing the form of some
constituent matrices, without altering the system parameters.
This change has marginal effects on the complexity of the cryptosystem that, 
instead, preserves its security against all known attacks.
This work gives an overview of the QC-LDPC codes-based McEliece cryptosystem
and its cryptanalysis.
Two recent versions are considered, and their ability to counter all 
the currently known attacks is discussed. A third version able to reach
a higher security level is also proposed.
Finally, it is shown that the new QC-LDPC codes-based cryptosystem scales
favorably when larger keys are needed, as very recently pointed out by
the successful implementation of an attack against the original 
cryptosystem.
\end{abstract}

\end{frontmatter}

\thispagestyle{empty}
\pagestyle{empty}

\section*{Introduction}

First presented by Robert J. McEliece in 1978 \cite{McEliece1978}, the McEliece cryptosystem 
represents one of the most famous examples of error correcting codes-based public key cryptosystem.
It adopts generator matrices of linear block codes as private and public keys, and the combination 
of a dense transformation and a permutation to hide the structure of the secret code into 
the public generator matrix.
Its security lies in the difficulty of decoding a large linear code having
no visible structure, that is an NP complete problem \cite{Berlekamp1978}.
The McEliece cryptosystem has successfully resisted cryptanalysis for thirty years, and no
algorithm able to realize a total break in a reasonable time has been found up to now.

Attacks achieving the lowest work factors aim at solving the general decoding problem, that
consists in deriving the error vector affecting a codeword of an $(n, k)$-linear block code
(\emph{i.e.}, having length $n$ and dimension $k$). It can be shown that this problem can be translated
into that of finding the minimum weight codeword in an $(n, k+1)$-linear block code, so
the McEliece cryptosystem can also be attacked by means of algorithms aimed at finding low 
weight codewords.

A first decoding attack was already proposed by McEliece in his paper \cite{McEliece1978}
and is based on the principle of \emph{information set decoding}. It consists in selecting $k$ bits 
of the ciphertext and inverting the encoding map, hoping that none of them is in error.
This attack has been further improved by Lee and Brickell \cite{Lee1988}, who proposed a systematic
procedure for validating the decoded words and showed that the attack can be attempted also 
when the chosen information set is affected by a small number of errors.

More recent decoding attacks are instead based on probabilistic algorithms searching for
low weight codewords. Stern's algorithm \cite{Stern1989} is among the most famous ones, 
and it has been later improved by Canteaut and Chabaud \cite{Canteaut1998}.
Very recently, Bernstein et al. have proposed a highly efficient implementation of the
attack based on Stern's algorithm \cite{Bernstein2008}, that is able to achieve a speedup
of about 12.
The improved algorithm has been run on a computer cluster, and an encrypted codeword of
the original McEliece cryptosystem has  been correctly deciphered, thus proving the 
feasibility of an attack for the original choice of the system parameters.

Despite this, no polynomial time attack has been found up to now, and the system remains 
secure, provided that large enough keys are adopted in order to reach suitable work factors
on modern computers.
In addition, the McEliece cryptosystem can be considered to be a \emph{post-quantum} cryptographic
system \cite{Bernstein2009}, since no polynomial time algorithm able to exploit quantum 
computers for an attack has been found up to now. On the contrary, Shor presented a quantum 
polynomial time algorithm for calculating discrete logarithms that should be able to break 
RSA, DSA and ECDSA \cite{Shor1997}.

Moreover, the original version of the McEliece cryptosystem, based on binary Goppa codes 
with irreducible generator polynomials, can be two or three orders of magnitude faster than 
RSA.
However, unlike RSA, the original McEliece cryptosystem has been rarely considered in 
practical applications, due to its two major drawbacks: large keys and low transmission rates.
Many attempts have been made for replacing Goppa codes with other families of codes in order 
to overcome such drawbacks, but they always compromised the system security.
This occurred for Generalized Reed-Solomon Codes \cite{Niederreiter1986} and Reed-Muller 
codes \cite{Sidelnikov1994}. Successful total break attacks have also been conceived for some
versions adopting Quasi-Cyclic (QC) codes \cite{Gaborit2005} and Low-Density Parity-Check 
(LDPC) codes \cite{Monico2000, Baldi2007ISIT}.

LDPC codes represent the state of the art in forward error correction and are able to
approach the ultimate capacity bounds \cite{Richardson2001}.
Their performance under belief propagation decoding depends on the characteristics of
their sparse parity-check matrices and their design can be performed on a random basis.
Thus, it is possible to obtain large families of equivalent codes, that is the first
requisite for their application in cryptography.
The adoption of LDPC codes in the McEliece cryptosystem can yield many advantages:
the sparse nature of their parity-check matrices could help to reduce the key size, 
at least in principle, and their easy design could allow to increase the transmission
rate.
Unfortunately, the usage of LDPC matrices as public keys can compromise the system 
security \cite{Monico2000, Baldi2007ISIT, Baldi2007ICC}.
For this reason, it has been proposed to adopt public keys in the form of
generator matrices of a particular family of QC-LDPC codes, that are structured
LDPC codes. Their structured character allows to reduce the key size though using
dense generator matrices.

Even with this choice, the adoption of sparse and block-wise diagonal transformation matrices
can still expose the cryptosystem to total break attacks \cite{Otmani2008}; so,
the original proposal has been recently revised in such a way to not include
this kind of matrices.
The new cryptosystem is immune against all currently know attacks, it allows
a significant reduction in the key size with respect to the original version 
and achieves increased transmission rate.
Furthermore, the size of its public keys increases linearly with the code dimension;
so the new cryptosystem scales favorably when larger keys are needed for facing
the growing computational power of modern computers.

The paper is organized as follows: Section \ref{sec:OriginalMcEliece} describes
the original McEliece cryptosystem, while Section \ref{sec:LDPCMcEliece} is focused 
on its variants based on LDPC codes. In Section \ref{sec:Attacks} the most dangerous 
attacks against the cryptosystem security are studied, together with their possible 
countermeasures.
Section \ref{sec:Complexity} is devoted to the complexity assessment of the considered
cryptosystems and Section \ref{sec:Conclusion} concludes the paper.

\section{The original McEliece cryptosystem}
\label{sec:OriginalMcEliece}

Inspired by the introduction of asymmetric cryptography by Diffie and Hellmann
\cite{Diffie1976}, McEliece proposed his code-based public key cryptosystem 
starting from the observation that a fast decoding algorithm exists for a
general Goppa code, while the same does not occur for a general linear code
\cite{McEliece1978}.

In the McEliece cryptosystem, Bob randomly chooses an irreducible polynomial 
of degree $t$ over $GF(2^m)$, that corresponds to an irreducible Goppa code 
of length $n=2^m$ and dimension $k \geq n-tm$, able to correct $t$ or fewer 
errors in each codeword.
Then, Bob produces a $k \times n$ generator matrix $\mathbf{G}$ for the
secret code, in reduced echelon form, that will be part of his secret key.
The remaining part of the secret key is formed by two other matrices:
a dense $k \times k$ non singular matrix $\mathbf{S}$ and a random $n \times n$ 
permutation matrix $\mathbf{P}$.

Then, Bob produces his public key as follows (the inverses of $\mathbf{S}$
and $\mathbf{P}$ are used here, rather than in the decryption map, for
consistency with the notation used for the new proposals):
\begin{equation}
\mathbf{G}' = \mathbf{S}^{-1} \cdot \mathbf{G} \cdot{\mathbf{P}^{-1}}.
\end{equation}

Alice, in order to send encrypted messages to Bob, fetches his public
key $\mathbf{G}'$ from the public directory, divides her message into $k$-bit
words, and applies the encryption map as follows:
\begin{equation}
\mathbf{x} = \mathbf{u} \cdot \mathbf{G}' + \mathbf{e},
\end{equation}
where $\mathbf{x}$ is the ciphertext corresponding to the cleartext $\mathbf{u}$
and $\mathbf{e}$ is a random vector of $t$ intentional errors.

\begin{figure}
\begin{centering}
\includegraphics[keepaspectratio, width=120mm]{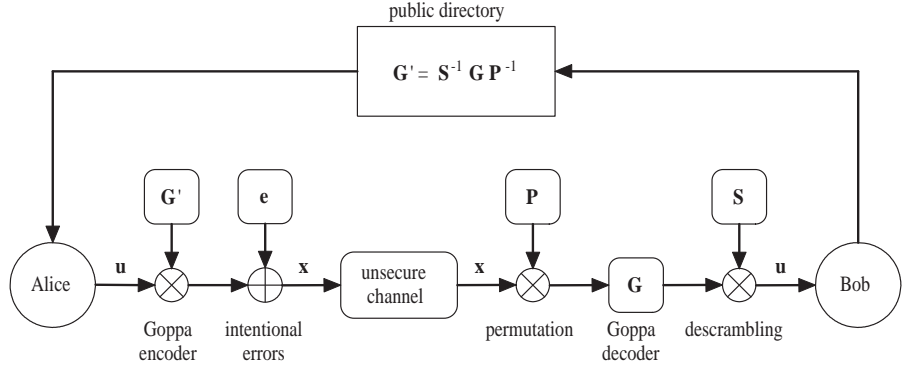}
\caption{The original McEliece cryptosystem.}\label{fig1}
\par\end{centering}
\end{figure}

Bob, after having received the encrypted message $\mathbf{x}$, inverts the secret
permutation, thus finding a codeword of the secret Goppa code affected by
the vector of intentional errors $\mathbf{e} \cdot \mathbf{P}$, having weight
$t$:
\begin{equation}
\mathbf{x}' = \mathbf{x} \cdot \mathbf{P} = \mathbf{u} \cdot \mathbf{S}^{-1} \cdot \mathbf{G} + \mathbf{e} \cdot \mathbf{P}.
\end{equation}
By exploiting Goppa decoding, Bob is able to correct all the $t$ intentional
errors. Hence he can obtain $\mathbf{u} \cdot \mathbf{S}^{-1}$, due to the
systematic form of $\mathbf{G}$, and then recover $\mathbf{u}$ through multiplication
by $\mathbf{S}$. The main blocks of the McEliece cryptosystem are shown
in Figure \ref{fig1}.

In his original formulation, McEliece adopted Goppa codes with length $n = 1024$
and dimension $k = 524$, able to correct up to $t = 50$ errors.
The key size is hence $n \times k = 67072$ bytes, and the transmission rate is
$k/n \approx 0.5$.
On the other hand, the RSA system with $1024$-bit modulus and public exponent
$17$ has keys of just $256$ bytes and reaches unitary transmission rate 
(\emph{i.e.}, encryption has no overhead on the transmission).

However, it must be considered that the McEliece cryptosystem is significantly
faster than RSA: it requires $514$ binary operations per bit for encoding and
$5140$ for decoding. On the contrary, RSA requires $2402$ and $738112$ binary 
operations per bit for encoding and decoding, respectively \cite{Canteaut1998}.

\section{LDPC codes in the McEliece cryptosystem}
\label{sec:LDPCMcEliece}

In this section a recent version of the McEliece cryptosystem based on QC-LDPC
codes is described.
It exploits the peculiarities of QC-LDPC codes for overcoming the drawbacks
of the original system and it is able to resist all attacks currently known.

First, some basic properties of QC-LDPC codes are reminded, then it is shown
how the McEliece cryptosystem should be modified in order to use these codes 
as private and public keys without incurring in security issues.

\subsection{QC-LDPC codes based on difference families}
\label{subsec:QCLDPConRDF}

LDPC codes represent a particular class of linear block codes, able to
approach channel capacity when soft decision decoding algorithms based
on the \emph{belief propagation} principle are adopted \cite{Richardson2001}.

An $(n, k)$ LDPC code $C$ is defined as the kernel of a sparse $(n-k) \times n$
parity-check matrix $\mathbf{H}$:
\begin{equation}
C = \left\{ \mathbf{c} \in GF(2)^n: \mathbf{H} \cdot \mathbf{c}^T = \mathbf{0} \right\}.
\end{equation}
In order to achieve very good performance under belief propagation decoding,
the parity-check matrix $\mathbf{H}$ must have a low density of 1 symbols
(typically on the order of $10^{-3}$) and absence of short cycles in the
associated Tanner graph. The shortest possible cycles, that have length four,
are avoided when any pair of rows (columns) has supports with no more than 
one overlapping position.

These conditions suffice to obtain good LDPC codes; so they can be designed
through algorithms that work directly on the parity-check matrix, aiming at
maximizing the cycles length, like the Progressive Edge Growth (PEG) 
algorithm \cite{Hu2005}.
The codes obtained are unstructured, in the sense that the positions of 1 symbols
in each row (or column) of the parity-check matrix are independent of the others.
This feature influences complexity of the encoding and decoding stages, since the
whole matrix must be stored and the codec implementation cannot take advantage
of any cyclic or polynomial nature of the code.
In this case, a common solution consists in adopting lower triangular or 
quasi-lower triangular parity-check matrices, that correspond to sparse
generator matrices, in such a way as to reduce complexity of the encoding
stage \cite{Richardson2001EfficientEncoding}.

Opposite to this approach, structured LDPC codes have also been proposed, whose
parity-check matrices have a very simple inner structure.
Among them, QC-LDPC codes represent a very important class,
able to join easy encoding of QC codes with the astonishing performance of
LDPC codes.
For this reason, QC-LDPC codes have been included in several recent
telecommunication standards and applications \cite{802.16e, CCSDS2007}.

QC-LDPC codes have both length and dimension multiple of an integer
$p$, that is, $n = n_0p$ and $k = k_0p$. They have the property
that each cyclic shift of a codeword by $n_0$ positions is still a
valid codeword. This reflects on their parity-check matrices, that
are formed by circulant blocks.
A $p \times p$ circulant matrix $\mathbf{A}$ over $GF(2)$ is defined
as follows:
\begin{equation}
\mathbf{A}=\left[{\begin{array}{ccccc}
{a_{0}} & {a_{1}} & {a_{2}} & \cdots & {a_{p-1}}\\
{a_{p-1}} & {a_{0}} & {a_{1}} & \cdots & {a_{p-2}}\\
{a_{p-2}} & {a_{p-1}} & {a_{0}} & \cdots & {a_{p-3}}\\
\vdots & \vdots & \vdots & \ddots & \vdots\\
{a_{1}} & {a_{2}} & {a_{3}} & \cdots & {a_{0}}\end{array}}\right],
\label{eq:CircMatrix}
\end{equation}
where $a_{i} \in GF(2), i = 0 \ldots p-1$.

A simple isomorphism exists between the algebra of $p\times p$ 
binary circulant matrices and the ring of polynomials $GF(2)[x]/(x^p+1)$.
If we denote by $\mathbf{X}$ the unitary cyclic permutation matrix,
the isomorphism maps $\mathbf{X}$ into the monomial $x$ and the
circulant matrix $\sum_{i=0}^{p-1}\alpha_i\mathbf{X}^i$ into the polynomial
$\sum_{i=0}^{p-1}\alpha_ix^i \in GF(2)[x]/(x^p+1)$.
This isomorphism can be easily extended to matrices formed by circulant blocks.

Let us focus attention on a particular family of QC-LDPC codes,
having the parity-check matrix formed by a single row of $n_0$
circulant blocks, each with row (column) weight $d_v$:
\begin{equation}
\mathbf{H} = \left[ \mathbf{H}_{0} | \mathbf{H}_{1} | \ldots |\mathbf{H}_{n_0-1} \right].
\label{eq:HCircRow}
\end{equation}
If we suppose (without loss of generality) that $\mathbf{H}_{n_0-1}$
is non singular, a valid generator matrix for the code in systematic form
can be expressed as follows:
\begin{equation}
\mathbf{G}=\left[\mathbf{I}\begin{array}{c}
\left(\mathbf{H}_{n_{0}-1}^{-1}\cdot\mathbf{H}_{0}\right)^{T}\\
\left(\mathbf{H}_{n_{0}-1}^{-1}\cdot\mathbf{H}_{1}\right)^{T}\\
\vdots\\
\left(\mathbf{H}_{n_{0}-1}^{-1}\cdot\mathbf{H}_{n_{0}-2}\right)^{T}\end{array}\right],
\end{equation}
where $\mathbf{I}$ represents the $k \times k$ identity matrix.

Very simple methods for designing parity-check matrices in
the form (\ref{eq:HCircRow}), free of length-4 cycles, are 
those exploiting differences families and their variants \cite{Johnson2003, Xia2005, Baldi2005DF}.
Such methods are based on the observation that, if we denote 
as $\mathbf{h}_i, i = 0 \ldots n_0-1$, the vector containing
the positions of 1 symbols in the first row of $\mathbf{H}_i$, 
the absence of length-4 cycles in $\mathbf{H}$ is ensured when 
all the $\mathbf{h}_i$'s have disjoint sets of differences modulo $p$.
Sets of $\mathbf{h}_i$'s with such property can be obtained on
a random basis, so yielding large families of codes with identical
parameters \cite{Baldi2007ISIT}.

All the codes in a family share the characteristics
that mostly influence performance of belief propagation decoding,
that are: code length and dimension, parity-check matrix density,
nodes degree distributions and cycles length distribution.
So, they have equivalent error correction performance under belief
propagation decoding.

\begin{figure}
\begin{centering}
\includegraphics{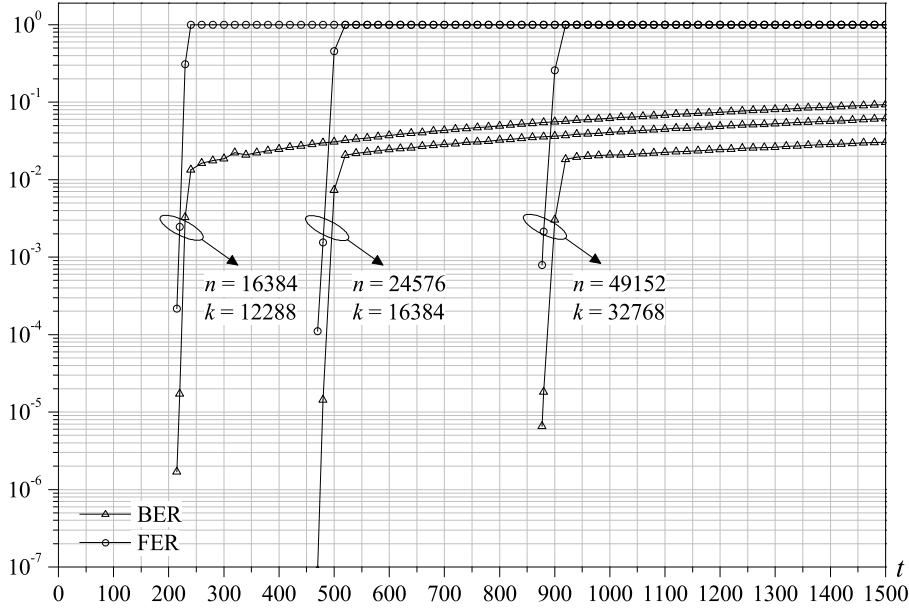}
\par\end{centering}
\caption{Performance of some QC-LDPC codes over the McEliece channel. 
\label{fig:Performance}}
\end{figure}

In order to apply such codes within the framework of the McEliece
cryptosystem, it is interesting to assess their error correction 
capability over a channel that adds exactly $t$ errors in each codeword.
This channel can be seen as a variant of the Binary Symmetric Channel (BSC),
and will be denoted as the \emph{McEliece channel} in the following.
This evaluation can be done through numerical simulations: Figure \ref{fig:Performance}
shows the performance in terms of Bit Error Rate (BER) and
Frame Error Rate (FER) of three QC-LDPC codes that will be of
interest in the following. They have $(n, k)=(16384, 12288)$,
$(24576, 16384)$ and $(49152, 32768)$, respectively.

It is important to note that the decoding radius of LDPC codes over
the McEliece channel cannot be determined analytically, as instead
occurs for Goppa codes; so, we can only choose values of $t$ that 
are able to ensure an extremely low error rate.

\subsection{McEliece cryptosystem adopting QC-LDPC codes}
\label{subsec:QCLDPCMcEliece}

The adoption of QC-LDPC codes in the McEliece cryptosystem can yield
important advantages in terms of key size and transmission rate.
As any other family of linear block codes, QC-LDPC codes are exposed
to the same attacks targeted to the original cryptosystem; among them,
decoding attacks represent the most dangerous ones (as it will be shown
in Section \ref{subsec:DecodingAttacks}).

Moreover, the adoption of LDPC codes could expose the system to
new attacks, due to the sparse nature of their matrices.
It was already observed in \cite{Monico2000} that LDPC matrices
cannot be used for obtaining the public key, not even after applying
a linear transformation through a sparse matrix.
In this case, the secret LDPC matrix could be recovered through
\emph{density reduction attacks}, that aim at finding the rows of
the secret matrix by exploiting their low density \cite{Monico2000, Baldi2006phd}.

One could think to replace LDPC matrices with their
corresponding generator matrices that, in general, are dense.
Actually, this is what happens in the original McEliece cryptosystem,
where a systematic generator matrix for the secret Goppa code
is used, hidden through a permutation.
However, a permutationally equivalent code of an LDPC code is
still an LDPC code, and the rows of its LDPC matrix could be
found by searching for low weight codewords in the dual of the
secret code. We call this strategy \emph{attack to the 
dual code}: it aims at finding a sparse representation for the
parity-check matrix of the public code, that can be used
for effective LDPC decoding.

So, when adopting LDPC codes in the McEliece cryptosystem, it
does not suffice to hide the secret code through a permutation,
but it must be ensured that the public code does not admit 
sparse characteristic matrices.
For this reason, it has been proposed to replace the permutation
matrix $\mathbf{P}$ with a different transformation matrix, $\mathbf{Q}$
\cite{Baldi2007ISIT}. $\mathbf{Q}$ is a sparse $n \times n$ matrix,
with rows and columns having Hamming weight $m>1$.
This way, the LDPC matrix of the secret code ($\mathbf{H}$) is 
mapped into a new parity-check matrix that is valid for the public code:
\begin{equation}
\mathbf{H}' = \mathbf{H} \cdot \mathbf{Q}^T.
\end{equation}
Depending on the value of $m$, the density of $\mathbf{H}'$ could
be rendered high enough to avoid attacks to the dual code.

In the modified cryptosystem, Bob chooses a secret LDPC code
by fixing its parity-check  matrix, $\mathbf{H}$, and selects
two other secret matrices: a $k \times k$ non singular scrambling
matrix $\mathbf{S}$ and an $n \times n$ non singular transformation 
matrix $\mathbf{Q}$ with row/column weight $m$.
Then, Bob obtains a systematic generator matrix $\mathbf{G}$ for
the secret code and produces his public key as follows:
\begin{equation}
\mathbf{G}' = \mathbf{S}^{-1} \cdot \mathbf{G} \cdot{\mathbf{Q}^{-1}}.
\end{equation}
It should be noted that the public key is a dense matrix,
so the sparse character of LDPC codes does not help reducing the
key length. However, when adopting QC-LDPC codes, the characteristic
matrices are formed by circulant blocks that are completely described 
by a single row or column.
This fact significantly reduces the key length that, moreover,
increases linearly with the code length.

The encryption map is the same as in the original cryptosystem:
$\mathbf{G}'$ is used for encoding and a vector $\mathbf{e}$ of
intentional errors is added to the encoded word.
The Hamming weight of vector $\mathbf{e}$, in this case, is
denoted as $t'$.
The decryption map must be slightly modified with respect to
the original cryptosystem.
After having received a ciphertext, Bob must invert the transformation
as follows:
\begin{equation}
\mathbf{x}' = \mathbf{x} \cdot \mathbf{Q} = \mathbf{u} \cdot \mathbf{S}^{-1} \cdot \mathbf{G} + \mathbf{e} \cdot \mathbf{Q},
\end{equation}
thus obtaining a codeword of the secret LDPC code affected by
the error vector $\mathbf{e} \cdot \mathbf{Q}$ with weight $\leq t=t'm$.
After that, Bob must be able to correct all the errors through
LDPC decoding and obtain $\mathbf{u} \cdot \mathbf{S}^{-1}$, due to the
systematic form of $\mathbf{G}$. Finally, he can recover $\mathbf{u}$ 
through multiplication by $\mathbf{S}$.

It should be noted that the introduction of the transformation matrix
$\mathbf{Q}$ in place of the permutation matrix causes an error amplification
effect (by a factor $m$). This is compensated by the error
correction capability of the secret LDPC code, that must be
able to correct $t$ errors. 

Based on this scheme, two possible choices of the system parameters
have been recently proposed, that are able to ensure different levels
of security against currently known attacks \cite{Baldi2008}.
A third choice is here considered that demonstrates how the cryptosystem
scales favorably when larger keys are needed for facing efficient
implementations of the attacks, as the one proposed recently.
For the three codes considered (whose performance is reported in Figure
\ref{fig:Performance}), $t=189$, $440$ and $780$ has been assumed, respectively,
and $m$ and $t'$ have been fixed accordingly.
The considered values of the parameters are summarized in Table \ref{tab:SysPars}.
It should be noted that the key size is simply $k_0n_0p$, since the whole
matrix can be described by storing only the first row (or column) of each 
circulant block.

\begin{table}
\begin{centering}
\caption{Choices of the parameters for the QC-LDPC-based McEliece
cryptosystem.}\label{tab:SysPars}
\begin{tabular}{ccccccc}
\hline
\textbf{System} & $n_0$ & $d_v$ & $p$ & $m$ & $t'$ & Key size (bytes) \\
\hline
1								& 4			& 13		& 4096	& 7		& 27 		& 6144 \\
\hline
2								& 3			& 13		& 8192	& 11	& 40		& 6144 \\
\hline
3								& 3			& 15		& 16384	& 13	& 60		& 12288 \\
\hline
\end{tabular}
\par\end{centering}
\end{table}

\section{Attacks and countermeasures}
\label{sec:Attacks}

For the sake of conciseness, this section considers only the attacks 
that are able to achieve the lowest work factors for the considered
cryptosystem, together with their possible countermeasures.

\subsection{Attacks to the dual code \label{subsec:DualCode}}

This kind of attacks exploits the fact that the dual of the public
code, that is generated by $\mathbf{H}'$, may contain low weight codewords,
and such codewords can be searched through probabilistic algorithms.
Each row of $\mathbf{H}'$ is a valid codeword of the dual code, so
it has at least $A_w \ge (n-k)$ codewords with weight $w \le d_{c}m$,
where $d_c = n_0d_{v}$ is the row weight of $\mathbf{H}$.

It should be observed that $d_{c} \ll n$ and the supports
of sparse vectors have very small (or null) intersection. So, by introducing 
an approximation, we can consider $A_w \approx (n-k)$.
With similar arguments, and assuming a small $m$, we can
say that the rows of $\mathbf{H}'$ have weight $w \approx d_{c}m = n_0d_vm$.

One of the most famous probabilistic algorithms for finding low 
weight codewords is due to Stern \cite{Stern1989} and exploits an 
iterative procedure.
When Stern's algorithm is performed on a code having length $n_S$
and dimension $k_S$, the probability of finding, in one iteration, one of $A_w$
codewords with weight $w$ is \cite{Hirotomo2005}:
\begin{equation}
P_{w,A_w} \leq A_{w} \cdot \frac{{w \choose g}{n_S-w \choose k_S/2-g}}{{n_S \choose k_S/2}} \cdot \frac{{w-g \choose g}{n_S-k_S/2-w+g \choose k_S/2-g}}{{n_S-k_S/2 \choose k_S/2}} \cdot \frac{{n_S-k_S-w+2g \choose l}}{{n_S-k_S \choose l}},
\label{eq:SternPw}
\end{equation}
where $g$ and $l$ are two parameters whose values must 
be optimized as functions of the total number of binary
operations. 
So, the average number of iterations needed to find a low weight
codeword is $c\geq P_{w,A_{w}}^{-1}$. Each iteration requires:
\begin{equation}
N=\frac{(n_S-k_S)^{3}}{2}+k_S(n_S-k_S)^{2}+2gl{k_S/2 \choose g}+\frac{2g(n_S-k_S){k_S/2 \choose g}^{2}}{2^{l}}
\label{eq:SternN}
\end{equation}
binary operations, so the total work factor is $\mathrm{WF}=cN$.

In the present case, Stern's algorithm is used for attacking the dual
of the public code, so $n_S = n$ and $k_S = n-k$.
Table \ref{tab:WFDualCode} reports the values of the maximum work factor
achieved (\emph{i.e.}, when $w = d_{c}m$) by the considered solutions,
together with the minimum value of $w$ needed to have work
factor $\ge 2^{80}$ (noted by $w(\mathrm{WF} \ge 2^{80})$ in the figure). 
Based on these results, it seems that all the three
systems can be considered secure against attacks to the dual code.

\begin{table}
\begin{centering}
\caption{Work factors of attacks to the dual code.}\label{tab:WFDualCode}
\begin{tabular}{ccccccc}
\hline
\textbf{System} & $n_0$ & $d_v$ & $p$ & $m$ & Max $\mathrm{WF}$ & $w(\mathrm{WF} \ge 2^{80})$ \\
\hline
1								& 4			& 13		& 4096	& 7		& $2^{153}$ & 179   \\
\hline
2								& 3			& 13		& 8192	& 11	& $2^{250}$ & 127  \\
\hline
3								& 3			& 15		& 16384	& 13	& $2^{340}$ & 124  \\
\hline
\end{tabular}
\par\end{centering}
\end{table}

\subsection{OTD attacks}
\label{subsec:OTDAttacks}

In the cryptosystem version proposed in \cite{Baldi2007ISIT}, both 
$\mathbf{S}$ and $\mathbf{Q}$ were chosen sparse, with non-null blocks 
having row/column weight $m$, and 
\begin{equation}
\mathbf{Q}=\left[\begin{array}{cc@{\ }c@{}c}
\ \mathbf{Q}_{0} & \mathbf{0} & \mathbf{0} & \mathbf{0}\quad\\
\mathbf{0} & \ \mathbf{Q}_{1} & \mathbf{0} & \mathbf{0}\quad\\
\mathbf{0} & \mathbf{0} & \ddots & \mathbf{0}\quad\\
\mathbf{0} & \mathbf{0} & \mathbf{0} & \quad\mathbf{Q}_{n_{0}-1}
\end{array}\right].
\end{equation}
This gave raise to an attack formulated by Otmani, Tillich and
Dallot, that is here denoted as \emph{OTD attack} \cite{Otmani2008}.

The rationale of this attack lies in the observation that, by selecting 
the first $k$ columns of $\mathbf{G}'$, an eavesdropper can obtain
\begin{equation}
\mathbf{G}'_{\leq k}=\mathbf{S}^{-1}\cdot\left[\begin{array}{@{}c@{}c@{\ \ }c@{\!}c}
\ \ \mathbf{Q}_{0}^{-1} & \mathbf{0} & \ldots & \mathbf{0}\\
\mathbf{0} & \ \ \mathbf{Q}_{1}^{-1} & \ldots & \mathbf{0}\\
\vdots & \vdots & \ddots & \vdots\\
\mathbf{0} & \mathbf{0} & \ldots & \quad\mathbf{Q}_{n_{0}-2}^{-1}\end{array}\right].
\end{equation}
Then, by inverting $\mathbf{G}'_{\leq k}$ and considering its block at position $(i,j)$, 
he can obtain $\mathbf{Q}_{i}\mathbf{S}_{i,j}$, that corresponds to the polynomial 
\begin{equation}
g_{i,j}(x) = q_{i}(x) \cdot s_{i,j}(x) \ \mathrm{mod}\left(x^{p}+1\right).
\end{equation}
If both $\mathbf{Q}_{i}$ and $\mathbf{S}_{i,j}$ are sparse, it is highly probable 
that $g_{i,j}(x)$ has exactly $m^2$ non-null coefficients and that its support contains 
at least one shift $x^{l_a} \cdot q_i(x)$, $0 \leq l_a \leq p-1$.

Three possible strategies have been proposed for implementing this attack.
According to the first strategy, the attacker can enumerate all the $m$-tuples
belonging to the support of $g_{i,j}(x)$. Each $m$-tuple can be then validated 
through inversion of its corresponding polynomial and multiplication by $g_{i,j}(x)$.
If the resulting polynomial has exactly $m$ non-null coefficients, the $m$-tuple 
is a shifted version of $q_i(x)$ with very high probability.
The second strategy exploits the fact that it is highly probable that the Hadamard 
product of the polynomial $g_{i,j}(x)$ with a $d$-shifted version of itself, 
$g_{i,j}^d(x) \ast g_{i,j}(x)$, gives a shifted version of $q_i(x)$, for a specific 
value of $d$.
The eavesdropper can hence calculate all the possible $g_{i,j}^d(x) \ast g_{i,j}(x)$
and check whether the resulting polynomial has $m$ non null coefficients.
As a third strategy, the attacker can consider the $i$-th row of the inverse of $\mathbf{G}'_{\leq k}$:
\begin{equation}
\mathbf{R}_{i}=\left[\mathbf{Q}_{i}\mathbf{S}_{i,0}|\mathbf{Q}_{i}\mathbf{S}_{i,1}|
\ldots|\mathbf{Q}_{i}\mathbf{S}_{i,n_{0}-2}\right].
\end{equation}
The linear code generated by
\begin{equation}
\mathbf{G}_{OTD3}=\left(\mathbf{Q}_{i}\mathbf{S}_{i,0}\right)^{-1} \cdot \mathbf{R}_{i} =
\left[\mathbf{I}|\mathbf{S}_{i,0}^{-1}\mathbf{S}_{i,1}|\ldots|\mathbf{S}_{i,0}^{-1}\mathbf{S}_{i,n_{0}-2}\right]
\end{equation}
admits an alternative generator matrix:
\begin{equation}
\mathbf{G}_{OTD3}'=\mathbf{S}_{i,0}\mathbf{G}_{OTD3}=
\left[\mathbf{S}_{i,0}|\mathbf{S}_{i,1}|\ldots|\mathbf{S}_{i,n_{0}-2}\right]
\end{equation}
that coincides with a block row of matrix $\mathbf{S}$.
When matrix $\mathbf{S}$ is sparse, the code defined by $\mathbf{G}_{OTD3}'$
contains low weight codewords.
Such codewords coincide with the rows of $\mathbf{G}_{OTD3}'$ and can be effectively 
searched through Stern's algorithm. 

With the choice of the parameters made in \cite{Baldi2007ISIT}, that
is almost coincident with the first choice in Table \ref{tab:SysPars},
the three OTD attack strategies would require, respectively, $2^{50.3}$,
$2^{36}$ and $2^{32}$ binary operations.
These low values can be easily reached with a standard computer, so
that cryptosystem must be considered broken.

However, the OTD attacks rely on the fact that both $\mathbf{S}$ and $\mathbf{Q}$ 
are sparse and that $\mathbf{Q}$ has block-diagonal form.
So, they can be effectively countered by adopting dense $\mathbf{S}$ matrices, 
without altering the remaining system parameters.
With dense $\mathbf{S}$ matrices the eavesdropper cannot obtain $\mathbf{Q}_{i}$ 
and $\mathbf{S}_{i,j}$, even knowing $\mathbf{Q}_{i}\mathbf{S}_{i,j}$,
the probability that the support of $g_{i,j}(x)$ contains that of at least one 
shift of $q_i(x)$ becomes extremely small and the code generated by $\mathbf{G}_{OTD3}$ 
does not contain any more low weight codewords.

For preserving the ability of correcting all the intentional errors, it is
important that $\mathbf{Q}$ remains sparse (with row/column weight $m$).
The choice of a dense $\mathbf{S}$ influences complexity of the decoding stage, 
that, however, can be reduced by resorting to efficient computation algorithms for circulant matrices \cite{Baldi2008}.

\subsection{Decoding attacks}
\label{subsec:DecodingAttacks}

As stated in the Introduction, the most promising attacks against the McEliece
cryptosystem are those aiming at solving the general decoding problem, that
is to obtain the error vector $\mathbf{e}$ used for encrypting a ciphertext. 

It can be easily shown that $\mathbf{e}$ can be searched as the lowest weight 
codeword in the extended code generated by 
\begin{equation}
\mathbf{G}''=\left[\begin{array}{c}\mathbf{G}'\\\mathbf{x}\end{array}\right].
\end{equation}

In order to evaluate the work factor of such attacks, we refer to Stern's 
algorithm, whose complexity can be easily evaluated in closed form, as already
shown in Section \ref{subsec:DualCode}.
Stern's algorithm has been further improved in \cite{Canteaut1998} and,
very recently, in \cite{Bernstein2008}.
Estimating the work factor of such modified algorithms is more involved, and requires
modeling the attack through Markov chains. For this reason, we continue to
refer to Stern's original formulation. For our purposes, it seems sufficient
to take into consideration that the adoption of optimized algorithms could 
result in a further speedup of about 12 times, as reported in \cite{Bernstein2008}.
According with the expressions reported in Section \ref{subsec:DualCode},
the work factor of a decoding attack against the original McEliece cryptosystem
based on Stern's algorithm would be $2^{63.5}$.

\begin{figure}
\begin{centering}
\includegraphics[keepaspectratio]{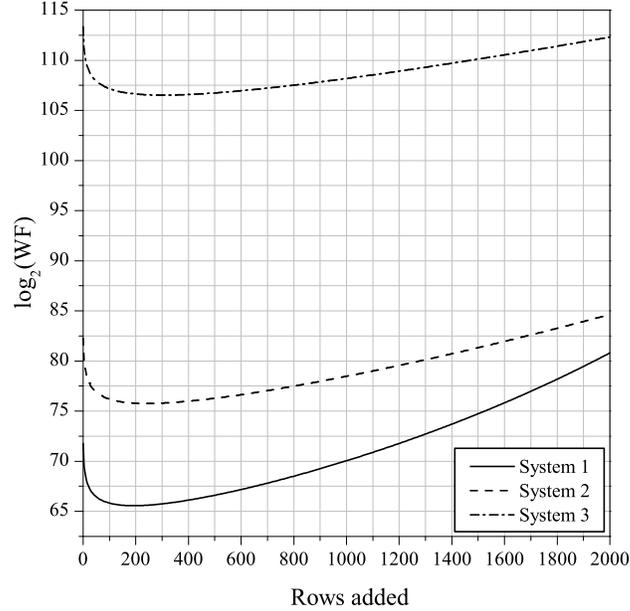}
\caption{Work factor of decoding attacks based on Stern's algorithm.}\label{fig:DecodingAttacks}
\par\end{centering}
\end{figure}

In the considered cryptosystem based on QC-LDPC codes, an extra speedup 
could result by considering the quasi-cyclic nature of the codes.
This yields that every block-wise cyclically shifted version of the ciphertext 
$\mathbf{x}$ is still a valid ciphertext.
So, an eavesdropper could continue extending $\mathbf{G}''$ by adding shifted
versions of $\mathbf{x}$, and could search for as many shifted versions of the
error vector.
Figure \ref{fig:DecodingAttacks} reports the values of the work factor of decoding 
attacks to the considered cryptosystem as functions of the number of rows added 
to $\mathbf{G}'$. The three considered choices of the system parameters reach, 
respectively, a minimum work factor of $2^{65.6}$, $2^{75.8}$ and $2^{106.5}$ 
binary operations.

Being the smallest work factors reached by currently known attacks, these values
can be considered as the security levels of the three cryptosystems.

\section{Complexity}
\label{sec:Complexity}

In order to compare the considered cryptosystems with more consolidated solutions, it is
important to estimate the complexity of both its encryption and decryption stages.

The encryption complexity is dominated by LDPC encoding, that coincides with calculating
the product $\mathbf{u}\cdot\mathbf{G}'$. The number of binary operations needed
by such task is denoted as $C_{mul}\left(\mathbf{u}\cdot\mathbf{G}'\right)$. Further $n$
operations must be considered for addition of the intentional error vector $\mathbf{e}$.
So, the encryption complexity can be expressed as follows:
\begin{equation}
C_{enc}=C_{mul}\left(\mathbf{u}\cdot\mathbf{G}'\right)+n.
\label{eq:Cenc}
\end{equation}
			
The computational cost of matrix multiplication can be reduced by exploiting the fact
that each matrix is formed by $p\times p$ binary circulant blocks.
Due to the isomorphism with the ring of polynomials $GF(2)[x]/(x^p+1)$, efficient algorithms 
for polynomial multiplication over finite fields can be adopted.
We refer to the Toom-Cook method, that is very efficient in the cases of our interest,
but other strategies are possible \cite{Baldi2008}.
	
As regards decryption complexity, it can be split into three contributions, corresponding
to: i) calculating the product $\mathbf{x}\cdot\mathbf{Q}$, ii) decoding the secret LDPC code
and iii) calculating the product $\mathbf{u}'\cdot\mathbf{S}$.
So, it can be expressed as follows:
\begin{equation}
C_{dec}=C_{mul}\left(\mathbf{x}\cdot\mathbf{Q}\right)+
C_{SPA}+C_{mul}\left(\mathbf{u}'\cdot\mathbf{S}\right)\label{eq:DecComplexity},
\label{eq:Cdec}
\end{equation}
where $C_{SPA}$ is the number of operations required for LDPC decoding through the sum-product 
algorithm.
By referring to the implementation proposed in \cite{Hu2001}, we can express $C_{SPA}$ as
follows:
\begin{equation}
C_{SPA}=I_{ave}\cdot n\left[q\left(8d_{v}+12R-11\right)+d_{v}\right],
\end{equation}
where $I_{ave}$ is the average number of decoding iterations and $q$ is the number of quantization 
bits used inside the decoder (both of them can be estimated through simulations).

By using Eq. (\ref{eq:Cenc}) and (\ref{eq:Cdec}), it is possible to estimate the encryption
and decryption cost in terms of binary operations per information bit.
This has been done in Table \ref{tab:Comparison}, that summarizes the main parameters 
of the considered cryptosystems and compares them with those of more consolidated solutions
(for the first three systems the complexity estimates are reported from \cite{Canteaut1998}).

\begin{table}
\begin{centering}
\caption{Parameters of the considered cryptosystems.}\label{tab:Comparison}
\begin{minipage}{\textwidth}
\begin{tabular}{ccccccc}
\hline
								&McEliece			&Niederreiter	&RSA 						&QC-LDPC		&QC-LDPC		&QC-LDPC\\
								&$(1024, 524)$&$(1024, 524)$&$1024$-bit mod.&McEliece 1	&McEliece 2	&McEliece 3\\
								&							&							&public exp. 17	&						&						&\\ 
\hline
Key	Size \footnote[1]{Expressed in bytes.} 
								&67072 				&32750 				&256   					&6144 			&6144				&12288\\
\hline
Rate						&0.51					&0.57					&1     					&0.75 			&0.67				&0.67\\
\hline
$k$ \footnote[2]{Information block length (bits).} 						
								&524   				&284   				&1024  					&12288			&16384			&32768\\
\hline
$C_{enc}/k$ \footnote[3]{Number of binary operations per information bit for encryption.}
								&514   				&50    				&2402  					&658  			&776				&1070\\
\hline
$C_{dec}/k$	\footnote[4]{Number of binary operations per information bit for decryption.}
								&5140  				&7863  				&738112					&4678 			&8901				&12903\\
\hline
\end{tabular}
\end{minipage}
\par\end{centering}
\end{table}

It can be noticed that all the three systems based on QC-LDPC codes have shorter keys and
higher rates with respect to the original McEliece cryptosystem and the Niederreiter version; so,
they succeed in improving their main drawbacks.
In particular, the first QC-LDPC-based system, that reaches a security level comparable
with that of the original McEliece cryptosystem, has key size reduced by more than 10 times 
with respect to it and more than 5 times with respect to the Niederreiter version.
Furthermore, the new system has increased transmission rate (up to $3/4$).

The security level can be increased at the expenses of the transmission rate: the second
QC-LDPC-based system has same key size as the first one, but its transmission rate
is reduced from $3/4$ to $2/3$. As a counterpart, its security level is increased by a
factor of about $2^{10}$.

Larger keys can be adopted in order to reach higher security levels, that are needed for 
facing efficient decoding attacks implemented on modern computers.
The third QC-LDPC-based system is able to reach a security level of $2^{106.5}$ by doubling the key
size (that is still more than 5 times smaller than in the original cryptosystem).
It should be noted that the system scales favorably when larger keys are needed, since
the key size grows linearly with the code length, due to the quasi-cyclic nature
of the codes, while in the original system it grows quadratically.

As concerns complexity, it can be observed that the first QC-LDPC-based cryptosystem
has encryption and decryption costs comparable with those of the original McEliece
cryptosystem. The Niederreiter version is instead able to significantly reduce the
encryption cost.
Encryption and decryption complexity increases for the other two QC-LDPC-based 
variants, but it still remains considerably lower with respect to RSA.
On the other hand, RSA has the smallest keys and reaches unitary rate.

\section{Conclusion}
\label{sec:Conclusion}

It has been shown that the adoption of LDPC codes in the framework of the McEliece
cryptosystem can help overcoming its drawbacks, that are large keys and low transmission
rate.
However, such choice must be considered carefully, since the sparse nature of the 
characteristic matrices of LDPC codes can expose the system to classic as well as 
newly developed attacks.
In particular, the misuse of sparse transformation matrices can expose the system to 
total break attacks, able to recover the secret key with reasonable complexity.

The adoption of dense transformation matrices permits to avoid such attacks, and
the quasi-cyclic nature of the codes still allows to reduce the key size.
Furthermore, the McEliece cryptosystem based on QC-LDPC codes can exploit efficient
algorithms for polynomial multiplication over finite fields for encryption and
low complexity LDPC decoding algorithms for decryption, that reduce its
computational complexity.

For these reasons, it seems that the considered variants of the McEliece cryptosystem
can be seen as a trade-off between its original version and other widespread solutions, 
like RSA.

\section*{Acknowledgments}

The author wishes to thank Franco Chiaraluce for his contribution
and Raphael Overbeck for helpful discussion on attacks.

\bibliographystyle{unsrt}
\bibliography{Archive}

\end{document}